\documentclass[fleqn,10pt]{wlscirep}
\usepackage[utf8]{inputenc}
\usepackage[T1]{fontenc}
\usepackage[left]{lineno}
\title{Coherent Optical Communications Enhanced by Machine Intelligence}

\author[1,+]{Sanjaya Lohani}

\author[1,*]{Ryan T. Glasser}
\affil[1]{Tulane University, New Orleans, LA 70118, USA}
\affil[+]{slohani@tulane.edu}
\affil[*]{rglasser@tulane.edu}

\begin{abstract}
Uncertainty in discriminating between different received coherent signals is integral to the operation of many free-space optical communications protocols, and is often difficult when the receiver measures a weak signal. Here we design an optical communications scheme that uses balanced homodyne detection in combination with an unsupervised generative machine learning and convolutional neural network (CNN) system, and demonstrate its efficacy in a realistic simulated coherent quadrature phase shift keyed (QPSK) communications system. Additionally, we program the neural network system at the transmitter such that it autonomously learns to correct for the noise associated with a weak QPSK signal, which is shared with the network state of the receiver prior to the implementation of the communications. We find that the scheme significantly reduces the overall error probability of the communications system, achieving the classical optimal limit. This communications design is straightforward to build, implement, and scale. We anticipate that these results will allow for a significant enhancement of current classical and quantum coherent optical communications technologies.

\end{abstract}
\begin{document}

\flushbottom
\maketitle
%
%
\thispagestyle{empty}
\section*{Introduction}
Nonorthogonality makes coherent states ideal candidates for various quantum and classical optical technology systems, such as quantum networks\cite{loock_optical_2011} and quantum metrology\cite{wiseman_quantum_2009, armen_adaptive_2002}, as well as classical optical communications schemes\cite{grosshans_quantum_2003,betti_coherent_1995,grosshans_continuous_2002}. Additionally, coherent states exhibit high spectral efficiency\cite{winzer_high-spectral-efficiency_2012,gnauck_optical_2005}, particularly in quadrature phase shift keying (QPSK), and are more tolerant to losses and nonlinearities in the communications channel\cite{giovannetti_classical_2004}. However, the accurate discrimination of coherent states is always limited by a fundamental uncertainty\cite{helstrom_quantum_2012}. This uncertainty is the basis of quantum key distribution (QKD)\cite{grosshans_quantum_2003} and multiple other communication security designs\cite{scarani_security_2009,bennett_quantum_1992,huttner_quantum_1995}. While this is a benefit for quantum communications, the discrimination uncertainty is always a barrier for efficient classical communications, in particular where the receiver site detects weak, low signal-to-noise ratio (SNR) coherent states such as in deep space communications\cite{kaushal_optical_2017}. As such, an essential characteristic of these coherent communications schemes, in this case QPSK, is the capacity to efficiently discriminate the signals that are sent and detected at the receiver. Realistic coherent optical communications systems involve the propagation of such signals through various obstacles that may decrease the SNR\cite{ke_experimental_2018} and thus increase the bit error rate. As a result, the noisy received QPSK signals can significantly limit the ability to establish such communications in a real-world environment. Here we make use of generative machine learning techniques\cite{vincent_extracting_2008} in combination with a convolutional neural network\cite{lohani_use_2018} to design a communications system that is robust to signal fading (or weak coherent signals) and demonstrate its ability to significantly reduce the error probability in discriminating between received coherent QPSK signals in a realistic simulated communications setting.

Generative machine intelligence systems have recently been applied to a variety of research areas\cite{sanchez-lengeling_inverse_2018,jang_ep-2092:_2018,li_machine_2018,donahue_exploring_2018,torlai_neural-network_2018}. Here we implement an unsupervised denoising autoencoder\cite{vincent_extracting_2008} as the generative neural network (GNN), a concept which has been demonstrated to be useful in multiple applications\cite{gondara_medical_2016,fichou_powerful_2018,cheng_deep_2018}.  Recently, several research groups have shown the power of machine intelligence in the context of coherent optical communications\cite{wang_intelligent_2017,zhang_intelligent_2018,lohani_turbulence_2018,khan_joint_2017, fan_demodulator_2017,kulin_end--end_2017}.
Here we use a generative neural network, in combination with a convolutional neural network (CNN), to make clean reconstructions and accurate classifications of received QPSK signals even for  extremely low SNR scenarios, which greatly reduces the overall error probability of the communications design, achieving the classical optimal (i.e homdyne limit\cite{proakis_digital_2000}) or standard quantum limit (SQL). In previous optical communications techniques using machine learning, unknown received optical signals (including images) have been classified with high accuracy using only a CNN as the classifier. It is, however, desirable for schemes with only a CNN as the classifier to have a training set with a large number of known homodyne measurements, which is always unbounded. This limits the discrimination efficiency of the communications design with respect to random signal (amplitude) fading of coherent states. Additionally, such setups with only a CNN as a classifier are based on supervised learning that requires a labelling for each training homodyne measurement. In practice, it is hard to accurately label the homodyne outputs of a received QPSK signal, particularly for very low SNRs. The present unsupervised learning strategy does not require any labelling for the weak (low SNR) coherent states, which are autonomously fed to the networks and reconstructed at the GNN outputs. Afterwards, the generated keys are classified by a CNN purely trained with the homodyne data associated with high SNR QPSK signals (desired), which are easily distinguished and labelled, prior to the start of communications (i.e., pre-trained before messages are sent and received). These novel aspects of the current design pave the way towards the robust and realistic implementation of homodyne receivers in efficiently demodulating signals in the coherent free space optical communications regime. 
\begin{figure}[h!]
\centering
\includegraphics[width=\linewidth]{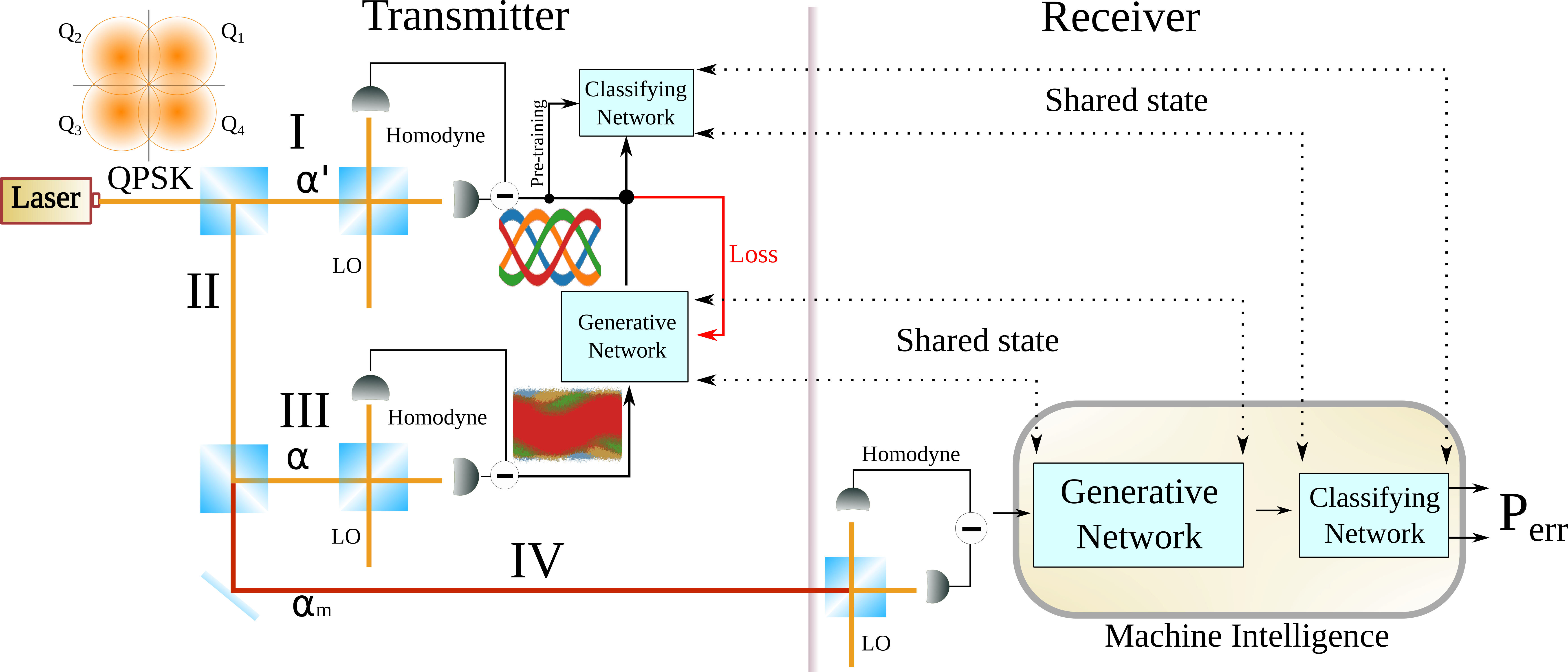}
\caption{Schematic of the robust coherent optical communications design with the machine intelligence aided homodyne receiver.  In the transmitter, QPSK signals ($Q_1,\,Q_2,\,Q_3,\,Q_4$) are simulated to propagate through paths I, III, and, IV. Prior to communication, the homodyne measurements of coherent states $|\alpha'\rangle$ along the path I are used as the target sets and training sets for the generative neural networks (GNN) and classifying convolutional neural networks (CNN), respectively. Similarly, the coherent states $|\alpha\rangle$ along the path III are used as the training sets for the generative network, whereas the coherent states $|\alpha_m\rangle$ along the path IV are the actual unknown message signals sent to and received at the receiver. Prior to communication, the neural networks setup at the receiver is shared with the neural networks state at the transmitter.  Then, weak unknown signals $|\alpha_m\rangle$ are sent along path IV and are efficiently demodulated at the homodyne receiver with the GNN and CNN. }
\label{fig:Figure_1}
\end{figure}

The overall communications system design is shown in Fig. \ref{fig:Figure_1}. A laser beam is split into two paths, path I and path II, with a beam splitter. The path II is further split into path III and IV. Path IV is the communications channel, and paths I -- III are used in training the networks prior to communication.  Note that the QPSK signals $|\alpha'\rangle$, and $|\alpha\rangle$ on path I and III, respectively, share the same QPSK state (key value) so that the homodyne detectors on both paths always read the same key value regardless the strength of noise (SNR) in either branch. The QPSK signal $|\alpha_{m}\rangle$ (``m'' for message) on the communications channel, path IV, is assumed to be an unknown message key that has been sent and received at the receiver end as shown in Fig. \ref{fig:Figure_1}. At the transmitter, the homodyne outputs of the QPSK signals $|\alpha\rangle$ on path III is fed to the input of the GNN, which reconstructs a new signal measurement at its output that is finally compared with the homodyne output of signals $|\alpha'\rangle$ from path I. Next we optimize the reconstruction loss and update the GNN parameter space (see ``Methods''). At the same time we use the homodyne measurements of the QPSK signals $|\alpha'\rangle$ to train the classifying CNN network (see ``Methods''). Note that the networks are pre-trained at the transmitter, which are assumed to be continuously shared with the networks at the receiver as indicated by dotted arrows in Fig. \ref{fig:Figure_1}. This pre-training may take place locally, then be distributed prior to implementation of the communications (which is the weak signal $|\alpha_{m}\rangle$ sent along path IV).  Finally, the machine intelligence aided homodyne receiver classifies unknown QPSK message signals $|\alpha_{m}\rangle$ at the receiver and the corresponding error probability is evaluated. Examples of the homodyne measurement of a received QPSK signal and corresponding clean reconstruction are shown in Fig. \ref{fig:Figure_2}. Furthermore, we evaluate the error probability for various combinations of $|\alpha\rangle$ (or $|\alpha_m\rangle$) and $|\alpha'\rangle$, various scanning phase ranges of the local oscillator, and various SNR values of the transmitted message signal $|\alpha_m\rangle$. Finally, we show a significant improvement in the overall error probability when the neural network system is used.  
\section*{Results}
 Here we simulate QPSK coherent signals $|\alpha'^k\rangle$, $|\alpha^k\rangle$  and $|\alpha_m^k\rangle$ for k $\in$ \{1,2,3,4\} with,
\begin{equation}
    \alpha'^k = |\alpha'|e^{i\phi^k}\, , \alpha^k = |\alpha|e^{i\phi^k},\quad \text{and}\, \alpha_m^k = |\alpha_m|e^{i\phi^k}\, \text{such that}\,\quad \phi^k\,=\,(k-\frac{1}{2})\frac{\pi}{2},
    \label{eqn1}
\end{equation}
where each coherent state has some mean amplitude (e.g. $|\alpha|$) and phase ($\phi$).  In order to simulate a balanced (one output from a 50:50 beam splitter is subtracted from the other as shown in Fig. 1) homodyne measurement, we use a local oscillator, $\beta\,=\,|\beta|e^{i\gamma}$, with amplitude $|\beta| \, >> |\alpha|$. Note that we set $\beta\,=\,100$ for all of the simulations presented in this paper. As a result, the mean signal, $\langle n \rangle$, and variance, $( \Delta n )^2$, at the detector are given by
\begin{equation}
    \langle n \rangle = 2|\beta|\langle \alpha| X_\gamma|\alpha \rangle\, \quad \text{and}\quad
     (\Delta n)^2\,=\,|\beta|^2; \quad \text{for} \quad X_\gamma = \frac{a^\dagger e^{i\gamma}+ae^{-i\gamma}}{2},\quad \text{and}
    \quad \gamma :\gamma + \pi/2
    \label{eqn2}
\end{equation}
where $a^\dagger$ and $a$ are the raising and lowering operators, respectively. 
Since the minimum error probability ($P_{err}$) of discriminating between the QPSK signals (keys) are bounded below classically by the homodyne limit\cite{proakis_digital_2000}, and quantum mechanically by the Helstrom limit\cite{noauthor_quantum_nodate-1}, we design and demonstrate that the present communication setups achieve the classical minimum error limit even for extremely low SNR scenarios. In order to calculate the SQL ($P^{HD}_{err}$) and the Helstrom limit ($P^{Hel}_{err}$), we use the following relations, which are discussed in detail in \cite{proakis_digital_2000,becerra_experimental_2013,izumi_displacement_2012},

\begin{equation}
    \begin{aligned}
    P^{HD}_{err}\,&=\,\textbf{erfc}(\frac{\alpha}{\sqrt{2}})\Big[1-\frac{1}{4}\textbf{erfc}(\frac{\alpha}{\sqrt{2}})\Big],\quad \text{and}\quad \\ 
    P^{Hel}_{err}\,&=\,1 - \frac{1}{8}e^{-|\alpha|^2}\Big(\sqrt{cosh|\alpha|^2 + cos|\alpha|^2} + \sqrt{sinh|\alpha|^2 + sin|\alpha|^2} + \sqrt{cosh|\alpha|^2 - cos|\alpha|^2} + \sqrt{sinh|\alpha|^2 - sin |\alpha|^2}\Big)^2,
    \end{aligned}
\end{equation}
where \textbf{erfc}(u)$\,=\,\frac{2}{\sqrt{\pi}}\int_{u}^{\infty}e^{-u^2}\,du$, is the complementary of the error function. 
\begin{figure}[h!]
\centering
\includegraphics[width=\linewidth]{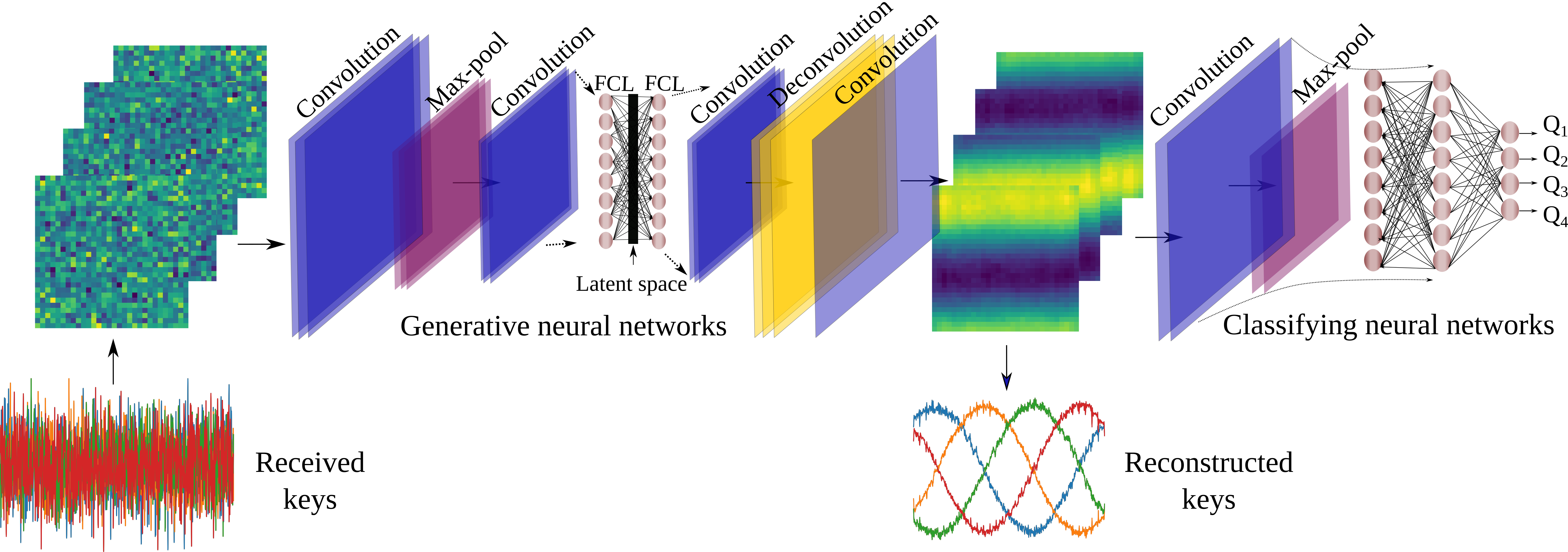}
\caption{Schematic of the architecture of the neural networks setup. Networks consist of generative neural networks (GNN) followed by classifying convolutional neural networks (CNN). The received QPSK homodyne signals are fed to the input of the GNN which reconstructs clean patterns as the output. The generated keys are then forwarded to the CNN, which classifies them. The abbreviation FCL stands for a fully connected layer.}
\label{fig:Figure_2}
\end{figure}

Next, we implement a GNN consisting of three sections -- an encoder, latent space, and a generator. The encoder learns the important features of the signals and stores them into a smaller dimension, which is a latent space. The latent space is then decoded through the generator which finally reconstructs the desired clean QPSK homodyne outputs. The implemented GNN uses convolutional units as the encoder and transpose-convolutional units as a decoder. The encoder contains two convolutional layers, a single max-pool layer, and a single fully connected layer (FCL) to the latent space, whereas the generator starts with a FCL followed by a convolutional layer, a transpose-convolutional layer, and again a convolutional layer as shown in Fig. \ref{fig:Figure_2}. Similarly, in order to classify the reconstructed, as well as uncorrected (noisy) QPSK homodyne outputs, we use a CNN at the end. The CNN consists of a single convolutional layer followed by max-pooling layer, and two FCLs. Finally, the FCL is connected with an output layer where classification decisions are made as shown in Fig. \ref{fig:Figure_2}. The architectures of the GNN and CNN are described in detail in the ``Methods'' section. The error probability introduced by the network is $P^{network}_{err}$, which is equal to the ratio of the total number of incorrect classifications to the total number of QPSK keys received, resulting in the overall error probability ($P_{err}$) in discriminating the received noisy QPSK signals at the receiver given by,
\begin{equation}
    P_{err}\,=\, 1 - (1-P^{HD}_{err})(1-P^{network}_{err}).
    \label{p_err}
\end{equation}
We then evaluate and plot the relative error probability with respect to the corresponding Helstrom limit such that relative Helstrom limit is scaled to 0 for all scenarios, which is given by,
\begin{equation}
     P^{relative}_{err} = P_{err} - P^{Hel}_{err},\, \quad \text{and}\, \quad P^{relative-HD}_{err} = P^{HD}_{err} - P^{Hel}_{err} .
    \label{relative_p_err}
\end{equation}
Throughout the rest of the manuscript, for convenience we plot/express both $P^{relative}_{err}$ and $P^{relative-HD}_{err}$ as the relative $P_{err}$, and a signal strength $|\alpha|$ in decibels, $|\alpha|\,dB\,=\,10\textrm{log}_{10}(|\alpha|)$.

\begin{figure}[h!]
\centering
\includegraphics[width=\linewidth]{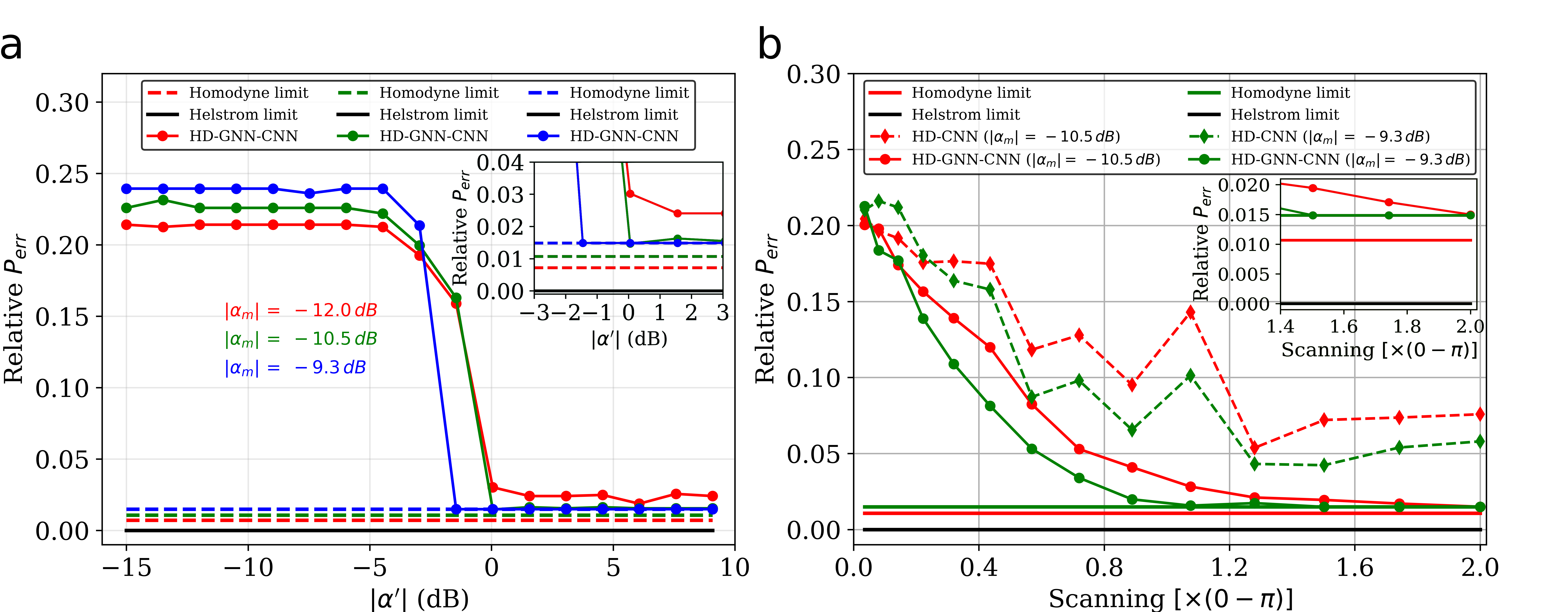}
\caption{(a) Relative error probability ($P_{err}$) versus target signal strength ($|\alpha'|$) at different SNR levels of message signal strength ($|\alpha_m|$), and (b) relative error probability versus scanning phase range of the local oscillator. The labels on the x-axis represent the phase range from 0 to the given value times $\pi$.  For example, the x-label 1.2 represents the phase range from 0 to 1.2$\pi$. HD-GNN-CNN shows the results from the networks setup that consists of both the GNN and classifying CNN at the receiver.} 
\label{fig:Figure_3}
\end{figure}

First we evaluate the relative $P_{err}$ as discussed in equation \ref{relative_p_err} with respect to the amplitude $|\alpha'|$ along the path I, given different SNRs in path III. Here we scan the local oscillator phase from 0 to $2\pi$ with a grid consisting of 900 points. We randomly simulate 200 homodyne outputs for each QPSK signal for 17 different values of $|\alpha'|$ from -15 dB to 9.08 dB, and set them as the target patterns at the output of the GNN. Similarly, we do the same for path III and randomly simulate 200 homodyne outputs for each QPSK signal for $|\alpha|$ = -12.0 dB, -10.5 dB, and -9.3 dB, and feed them to the encoder of GNN. Note that before feeding the signals to the GNN, we convert the homodyne outputs (from path I and III) into corresponding $30\times30$ pixels images as shown in Fig. \ref{fig:Figure_2}. In order to optimize $|\alpha'|$ at various values of $|\alpha|$, we keep $|\alpha|$ fixed at -12.0 dB, -10.5 dB, and -9.3 dB, and vary $|\alpha'|$ for all of them separately, such that 17 different GNN configurations are trained up to 150 epochs separately for each $|\alpha|$, for a total of 51 pre-trained GNNs. Also, we use the same $|\alpha'|$ data set, separately, to pre-train the classifying CNN networks at the end of the receiver. Next, we randomly simulate 360 homodyne outputs (90 per each QPSK key) for each $|\alpha_m|$ = -12.0 dB, -10.5 dB, and -9.3 dB as the test sets. Note that training data and test data share no similarity as they are randomly simulated at different times. Finally at the receiver, the homodyne measurements of $|\alpha_m|$ are fed to the shared pre-trained GNN which generates desired homodyne measurement as the output. Next, in order to calculate relative $P_{err}$, the reconstructed images (generated homodyne measurement outputs) from the GNN are forwarded to a pre-trained CNN and the corresponding relative $P_{err}$ is measured, the results of which are shown in Fig. \ref{fig:Figure_3} (a).  The relative $P_{err}$ for $|\alpha_m|$ = -12.0 dB, -10.5 dB and -9.3 dB at various $|\alpha'|$ are shown by the solid red, green, and blue curves, respectively, with the same color dotted lines representing the corresponding relative homodyne limit ($P^{relative-HD}_{err}$), and black solid line representing the relative Helstrom limit as shown in Fig. \ref{fig:Figure_3} (a). We find an improvement in relative $P_{err}$ begins when $|\alpha'|$ $\geq$ -4.5 dB for all $|\alpha_m|$ = -12.0 dB, -10.5 dB, and -9.3 dB.
As expected we obtain better reconstructions and less relative $P_{err}$ as we increase $|\alpha'|$ up to -1.5 dB, 3.1 dB, and 4.6 dB for $|\alpha_m|$ = -12.0 dB, -10.5 dB, and -9.3 dB, respectively, after which they begin to saturate. Additionally, we show that the system's relative $P_{err}$ achieves the corresponding homodyne limt for $|\alpha_m|$ = -9.3 dB at $|\alpha'|$ = -1.5 dB. Similarly we find a difference of $3.9\times10^{-3}$, and $1.1\times10^{-2}$, respectively, between the relative $P_{err}$ and corresponding homodyne limit for $|\alpha_m|$ = -10.5 dB, and -12.5 dB when $|\alpha'|$ = 0.05 dB, and 6.1 dB. Moreover, the corresponding relative $P_{err}$ for $|\alpha'|$ from -3 dB to 3 dB at the given $|\alpha_m|$ are zoomed in and shown in upper right inset of Fig. \ref{fig:Figure_3} (a). After taking into account the various SNR scenarios, we set $|\alpha'|$ relatively high at 9 dB as the optimized value for all simulations and results discussed in the following paragraphs. Note that this signal is used in pre-training the neural networks prior to communciation, and is not the signal that would be sent through a realistic communications channel (which is $|\alpha_{m}|$, and is simulated to be much weaker as discussed in the following).
\begin{figure}[b!]
\centering
\includegraphics[width=.7\linewidth]{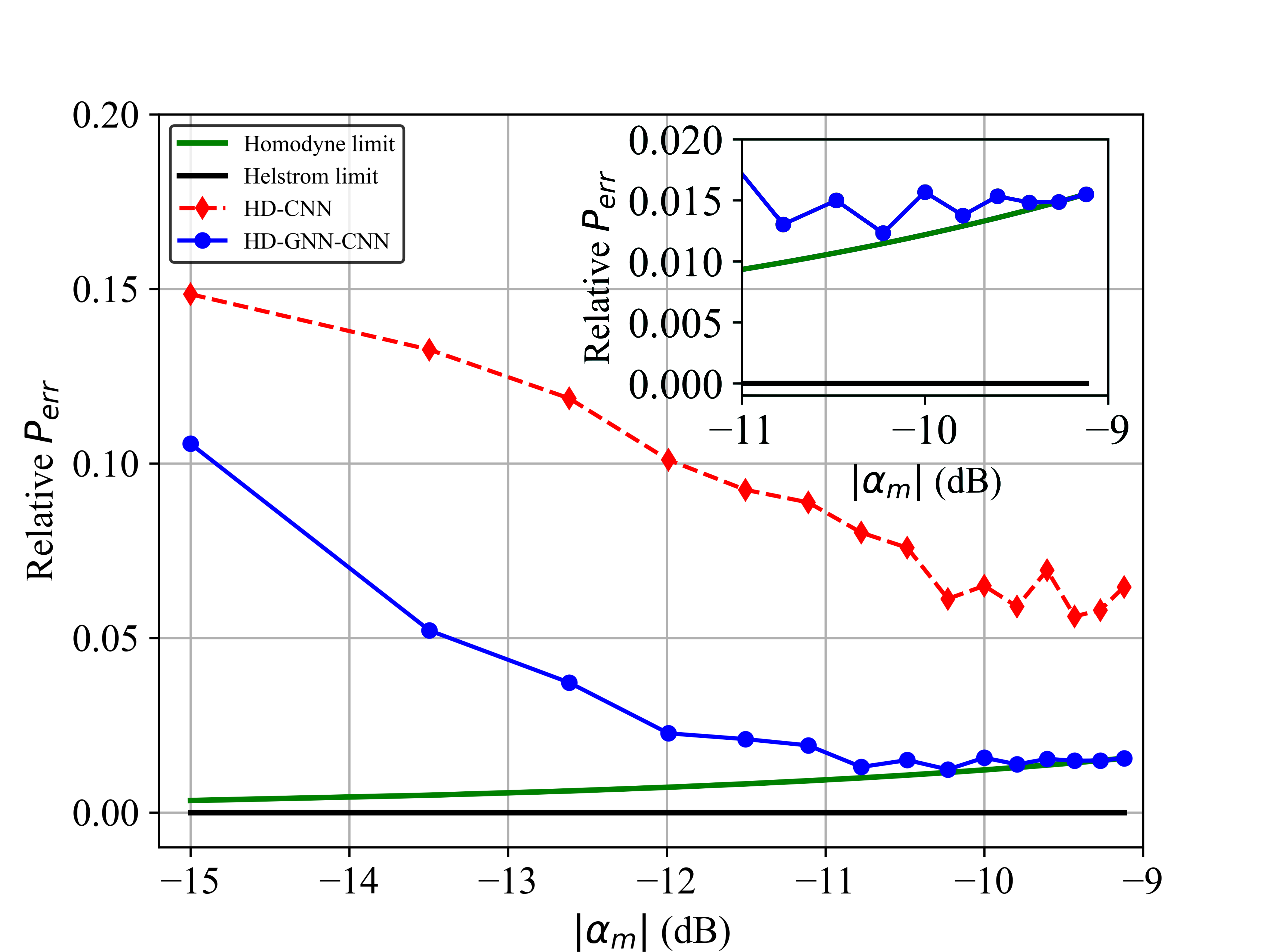}
\caption{Relative error probability ($P_{err}$) versus signal strength of the message signal ($|\alpha_m|$). The GNN is pre-trained with the targets of strength $|\alpha'|\,=\,9\,dB$. Similarly, the same data set for $|\alpha'|\,=\,9\,dB$ is used to pre-train the classifying CNN networks. The abbreviations HD-CNN and HD-GNN-CNN, respectively, show the results without and with GNN in the communications setup at the receiver.}
\label{fig:Figure_4}
\end{figure}
Next, we investigate the relative $P_{err}$ as the phase scanning range of the detection system is varied. We use the same training-test data and network settings as discussed above. In order to simulate training and test sets at various phase scanning ranges, we slice the grid of the homodyne outputs into a number of segments, for example as discussed in previous paragraph, into a grid of $30\times30$ (i.e. $0-900$ points) which represents a phase scanning from 0 to 2$\pi$. Here, we slice it into $28\times28$ ($0-784$ points), $26\times26$ ($0-676$ points), $24\times24$ ($0-576$ points) and so on up to $4\times4$ ($0-16$ points), for a total of 14 various scanning ranges.  These correspond to  phase scanning ranges from $0-1.74\pi$, $0-1.5\pi$, $0-1.28\pi$, and so on up to $0-0.04\pi$, respectively. Note that we separately train the GNN and CNN for each scanning range. Also we set the latent space dimension of the GNN to half of the input dimension for a given range.  For example, the input of $28\times28$ and $12\times12$ points, respectively, have latent space sizes of $14\times14$ and $6\times6$ points. With the GNN pre-trained for an $|\alpha|$ = -10.5 dB, and -9.3 dB, and the CNN pre-trained with $|\alpha'|$ = 9 dB, we calculate the relative $P_{err}$ for $|\alpha_m|$ = -10.5 dB, and -9,3 dB at the given scannning range. Here we reconstruct the 360 homodyne measurements (90 per each QPSK) for the given test set $|\alpha_m|$ and predict the corresponding QPSK value using the CNN, which is finally used to evaluate the relative $P_{err}$ . The relative $P_{err}$ resulting from the networks with and without the GNN versus scanning range is shown in Fig. \ref{fig:Figure_3} (b). The improvement in relative $P_{err}$ for $|\alpha_m|$ = -10.5 dB is shown by red curves, where the dotted red curve, solid red curve, and horizontal red line represent the relative $P_{err}$ without the GNN, with the GNN, and the relative homodyne limit, respectively. Similarly, the dotted green curve, solid green curve and a horizontal green line represent the relative $P_{err}$ without the GNN, with the GNN, and the relative homodyne limit for $|\alpha_m|$ = -9.3 dB, respectively. As discussed above, the black horizontal line is the relative Helstrom limit. We find a gradual improvement in the relative $P_{err}$ as we increase the scanning phase range of the local oscillator. Furthermore, with the aid of the GNN in the network and a scanning range of $0-0.87\pi$, we show significant improvement in relative $P_{err}$ from $9.5\times10^{-2}$ to $4.1\times10^{-2}$, and $6.5\times10^{-2}$ to $1.9\times10^{-2}$ for $|\alpha_m|$ = -10.5 dB, and -9.3 dB, respectively. Note that the relative homodyne limit for $|\alpha_m|$ = -9.3 dB, and -10.5 dB are $1.5\times10^{-2}$, and $1.1\times10^{-2}$, respectively. Additionally, we find the relative $P_{err}$ with the aid of the GNN is minimized and achieves the corresponding relative homodyne limit when the scanning range is greater than or equal to  $0-1.14\pi$ for $|\alpha_m|$ = -9.3 dB. A minimum difference of $4.3\times10^{-3}$ between the GNN-aided relative $P_{err}$ and relative homodyne limit is achieved for $|\alpha_m|$ = -10.5 dB at a scanning range of $0-2\pi$. The relative $P_{err}$ corresponding to a scanning range of $0-1.28\pi$, $0-1.74\pi$, and $0-2\pi$ are zoomed in and shown in upper right inset of Fig. \ref{fig:Figure_3} (b). 

We now turn to investigating the improvement of the relative $P_{err}$ at various SNR levels of transmitted message signals $|\alpha_m|$. In order to generate training sets for the GNN, we randomly simulate 200 homodyne measurements for each QPSK key for 15 values of $|\alpha|$ from -15.0 dB to -9.12 dB. Similarly, we randomly generate 90 homodyne measurements for each QPSK for 15 values of $|\alpha_m|$ from -15.0 dB to -9.12 dB as the test set separately. As a result, test sets and training sets again share no similarity. As mentioned earlier, we set $|\alpha'|$ to 9 dB, and the corresponding 200 homodyne outputs for each QPSK signal discussed above are used as the target for the GNN. The same $|\alpha'|$ is used to train the CNN. Note that we train separately the GNN for each value of $|\alpha|$, giving a total of 15 GNNs, while the classifying network (CNN) remains the same (as they share same target of the GNN). With the pre-trained GNN and CNN, we evaluate the relative $P_{err}$ with and without the GNN in the networks, results of which are shown in Fig. \ref{fig:Figure_4}. The dotted red and solid blue curves represent the relative $P_{err}$ without the GNN and with the GNN in the networks, respectively. Similarly, a solid green and a horizontal black line represent the relative homodyne and Helstrom limits for various values of $|\alpha_m|$. With the aid of the GNN in the networks, we achieve a remarkable improvement in relative $P_{err}$ from $6.1\times10^{-2}$ to $1.1\times10^{-2}$, which nearly reaches the relative optimal homodyne limit of $1.2\times10^{-2}$ for $|\alpha_m|$ = -10.23 dB. Similarly, for $|\alpha_m|$ = -9.27 dB, the relative $P_{err}$ achieves an improvement from $5.8\times10^{-2}$ to the optimal homodyne limit as shown in upper right inset of Fig. \ref{fig:Figure_4}. Furthermore, even for a very low SNR of $|\alpha_m|$ = -13.26 dB, we significantly reduce the difference between relative $P_{err}$ and the relative homodyne limit from $12.8\times10^{-2}$ to $4.7\times10^{-2}$.
\section*{Discussion}
In conclusion, we have designed a novel coherent optical communications setup that efficiently demodulates weak coherent QPSK states with a robust machine intelligence aided balanced homodyne receiver. The developed state-of-the-art GNN and CNN system, in combination, corrects for coherent QPSK signals associated with a wide range of SNRs, resulting in significantly reduced overall error probability of the communications system, either achieving or approaching the classical optimal limit. Additionally, by using the same network system, we also show an improvement in discrimination error probability for various scanning ranges of the local oscillator at different SNRs. Furthermore, with the aid of the GNN, which reconstructs a clean and desired quadrature measurement at its output, we have minimized the relative error probability with a CNN that is exclusively trained with homodyne measurements associated with high SNR (desired) coherent QPSK signals.  This allows bypassing the need for extremely large CNN training sets that would require various noises which are not only difficult to label, but are also unbounded. The present advances in QPSK demodulation and classification are essential to the robust performance of realistic coherent optical communications systems, and we anticipate that the presented techniques may directly be applied to an enhancement of current classical and quantum communications protocols.
\section*{Methods}
\subsection*{Architecture of the GNN}
 The GNN consists of three sections -- an encoder, latent space, and a decoder (generator). The encoder begins with a two dimensional convolutional layer with a kernel of size [5,5], stride length of 1, batch size of 10, feature mappings of 20, same padding, and a ReLU activation function. After this we apply a dropout with a rate of $20\%$ followed by a max-pool layer with a kernel of size [2,2] with a stride length of 2.  Then, again, we connect a two dimensional convolutional layer to the max-pool with the same parameters and dropout as discussed above. Next we attach a fully connected layer with a number of neurons that is always equal to the size of the latent space. Note that for the consistency we always choose the size of the latent space as half of the input dimension.  For example, input dimensions of $30\times30$, and $16\times16$, respectively, have a $15\times15$, and $8\times8$ sized latent space. Finally, the encoder convolutes the input dimension of \Big[10 (batch size), width, height, 1 (input channel)\Big] to \Big[10, latent space dimension\Big].  At the beginning of the decoder, the latent space is connected to a fully connected layer with \Big[width/2$\times$height/2$\times$20 (feature mappings)\Big] neurons, which is followed by a dropout with a rate of $20\%$. After this we apply a convolutional layer with stride length of 1 and transpose-convolutional (or deconvolutional) layer with stride length of 2, each with a kernel of size [5,5], feature mappings of 20, same padding, and a ReLU activation function. Note that each layer is followed by a dropout with a rate of $20\%$. Finally a convolutional unit with a single feature mapping generates a clean, or less noisy, homodyne measurement as the output. Here, the generator decodes the latent space of size \Big[10 (batch size), latent space dimension\Big] into the outputs of size \Big[10 (batch size), width, height\Big]. In order to optimize the clean reconstructions, a square reconstruction loss (error) \textbf{L}$\Big(GNN(\alpha),GNN(\alpha')\Big)$ is evaluated, where $GNN(\alpha)$ is the output of the GNN, and $GNN(\alpha')$ is the target. Finally, we minimize the average reconstruction loss given by equation (\ref{eqn:loss}) using adamoptimizer of tensorflow \cite{martin_abadi_tensorflow:_2015} with a learning rate of 0.008.
\begin{equation}
\theta\,,\theta^\prime = \underset{\theta\,,\theta^\prime}{\text{argmin}}\, \frac{1}{N}\sum_i^N \textbf{L}\Big(GNN(\alpha),GNN(\alpha')\Big),
\label{eqn:loss}
\end{equation}
where $\theta$ and $\theta'$ represent the encoder and decoder parameter space, respectively. The schematic of the GNN is shown in Fig. \ref{fig:Figure_2}.

\subsection*{Architecture of the classifying CNN}
The classifying CNN network begins with a convolutional unit with a kernel of size [2,2], batch size of 1, feature mappings of 10, same padding, stride length of 1, and a ReLU activation function. This is followed a max-pool layer with a kernel of size [2,2] and stride length of 2. Then we connect fully connected layers (FCLs) with 400 neurons and 50 neurons, which are consecutively followed by dropout layers with a rate of $80\%$, and $40\%$, respectively. Note that we use ReLU activations for both FCLs. Next we attach a final FCL with 4 neurons with a linear activation function to the end, followed by a softmax operation. In order to train a CNN to classify the generated clean homodyne outputs, we always use the target set of the GNN i.e. the homodyne measurements at the signal strength of $|\alpha'|$ as shown in path I of Fig. \ref{fig:Figure_1}. We randomly simulate 200 homodyne outputs per each QPSK key for the given $|\alpha'|$ as discussed in the results section, resulting in a total of 800 homodyne outputs. The data set is split into a training set with 170 measurements and a test set with 30 measurements, again, per each QPSK. The target of each QPSK is set as the one-hot vector output, for example, the first target key is [1,0,0,0], the second key is [0,1,0,0], and so on. After this the parameter space of the CNN is optimized by minimizing a softmax cross-entropy loss function using adamoptimizer with a learning rate of 0.001 up to 10 epochs. The neural networks' hyper-parameters are manually optimized as discussed in \cite{lohani_use_2018}. Note that pre-trained CNN network has unity accuracy with respect to the test homodyne measurements. The schematic of the classifying CNN is shown in Fig. \ref{fig:Figure_2}.

\bibliography{QPSK}

\begin{thebibliography}{10}
\urlstyle{rm}
\expandafter\ifx\csname url\endcsname\relax
  \def\url#1{\texttt{#1}}\fi
\expandafter\ifx\csname urlprefix\endcsname\relax\def\urlprefix{URL }\fi
\expandafter\ifx\csname doiprefix\endcsname\relax\def\doiprefix{DOI: }\fi
\providecommand{\bibinfo}[2]{#2}
\providecommand{\eprint}[2][]{\url{#2}}

\bibitem{loock_optical_2011}
\bibinfo{author}{Loock, P.~v.}
\newblock \bibinfo{journal}{\bibinfo{title}{Optical hybrid approaches to
  quantum information}}.
\newblock {\emph{\JournalTitle{Laser \& Photonics Reviews}}}
  \textbf{\bibinfo{volume}{5}}, \bibinfo{pages}{167--200},
  \doiprefix\url{10.1002/lpor.201000005} (\bibinfo{year}{2011}).

\bibitem{wiseman_quantum_2009}
\bibinfo{author}{Wiseman, H.~M.} \& \bibinfo{author}{Milburn, G.~J.}
\newblock \emph{\bibinfo{title}{Quantum {Measurement} and {Control}}}
  (\bibinfo{year}{2009}).

\bibitem{armen_adaptive_2002}
\bibinfo{author}{Armen, M.~A.}, \bibinfo{author}{Au, J.~K.},
  \bibinfo{author}{Stockton, J.~K.}, \bibinfo{author}{Doherty, A.~C.} \&
  \bibinfo{author}{Mabuchi, H.}
\newblock \bibinfo{journal}{\bibinfo{title}{Adaptive {Homodyne} {Measurement}
  of {Optical} {Phase}}}.
\newblock {\emph{\JournalTitle{Physical Review Letters}}}
  \textbf{\bibinfo{volume}{89}}, \bibinfo{pages}{133602},
  \doiprefix\url{10.1103/PhysRevLett.89.133602} (\bibinfo{year}{2002}).

\bibitem{grosshans_quantum_2003}
\bibinfo{author}{Grosshans, F.} \emph{et~al.}
\newblock \bibinfo{journal}{\bibinfo{title}{Quantum key distribution using
  gaussian-modulated coherent states}}.
\newblock {\emph{\JournalTitle{Nature}}} \textbf{\bibinfo{volume}{421}},
  \bibinfo{pages}{238}, \doiprefix\url{10.1038/nature01289}
  (\bibinfo{year}{2003}).

\bibitem{betti_coherent_1995}
\bibinfo{author}{Betti, S.}, \bibinfo{author}{Marchis, G.~D.} \&
  \bibinfo{author}{Iannone, E.}
\newblock \emph{\bibinfo{title}{Coherent {Optical} {Communications} {Systems}}}
  (\bibinfo{publisher}{Wiley-Interscience}, \bibinfo{address}{New York},
  \bibinfo{year}{1995}), \bibinfo{edition}{1 edition} edn.

\bibitem{grosshans_continuous_2002}
\bibinfo{author}{Grosshans, F.} \& \bibinfo{author}{Grangier, P.}
\newblock \bibinfo{journal}{\bibinfo{title}{Continuous {Variable} {Quantum}
  {Cryptography} {Using} {Coherent} {States}}}.
\newblock {\emph{\JournalTitle{Physical Review Letters}}}
  \textbf{\bibinfo{volume}{88}}, \bibinfo{pages}{057902},
  \doiprefix\url{10.1103/PhysRevLett.88.057902} (\bibinfo{year}{2002}).

\bibitem{winzer_high-spectral-efficiency_2012}
\bibinfo{author}{Winzer, P.~J.}
\newblock \bibinfo{journal}{\bibinfo{title}{High-{Spectral}-{Efficiency}
  {Optical} {Modulation} {Formats}}}.
\newblock {\emph{\JournalTitle{Journal of Lightwave Technology}}}
  \textbf{\bibinfo{volume}{30}}, \bibinfo{pages}{3824--3835}
  (\bibinfo{year}{2012}).

\bibitem{gnauck_optical_2005}
\bibinfo{author}{Gnauck, A.~H.} \& \bibinfo{author}{Winzer, P.~J.}
\newblock \bibinfo{journal}{\bibinfo{title}{Optical phase-shift-keyed
  transmission}}.
\newblock {\emph{\JournalTitle{Journal of Lightwave Technology}}}
  \textbf{\bibinfo{volume}{23}}, \bibinfo{pages}{115--130},
  \doiprefix\url{10.1109/JLT.2004.840357} (\bibinfo{year}{2005}).

\bibitem{giovannetti_classical_2004}
\bibinfo{author}{Giovannetti, V.} \emph{et~al.}
\newblock \bibinfo{journal}{\bibinfo{title}{Classical {Capacity} of the {Lossy}
  {Bosonic} {Channel}: {The} {Exact} {Solution}}}.
\newblock {\emph{\JournalTitle{Physical Review Letters}}}
  \textbf{\bibinfo{volume}{92}}, \bibinfo{pages}{027902},
  \doiprefix\url{10.1103/PhysRevLett.92.027902} (\bibinfo{year}{2004}).

\bibitem{helstrom_quantum_2012}
\bibinfo{editor}{Helstrom, C.~W.} (ed.) \emph{\bibinfo{title}{Quantum detection
  and estimation theory}} (\bibinfo{publisher}{Academic Press},
  \bibinfo{year}{2012}).

\bibitem{scarani_security_2009}
\bibinfo{author}{Scarani, V.} \emph{et~al.}
\newblock \bibinfo{journal}{\bibinfo{title}{The security of practical quantum
  key distribution}}.
\newblock {\emph{\JournalTitle{Reviews of Modern Physics}}}
  \textbf{\bibinfo{volume}{81}}, \bibinfo{pages}{1301--1350},
  \doiprefix\url{10.1103/RevModPhys.81.1301} (\bibinfo{year}{2009}).

\bibitem{bennett_quantum_1992}
\bibinfo{author}{Bennett, C.~H.}
\newblock \bibinfo{journal}{\bibinfo{title}{Quantum cryptography using any two
  nonorthogonal states}}.
\newblock {\emph{\JournalTitle{Physical Review Letters}}}
  \textbf{\bibinfo{volume}{68}}, \bibinfo{pages}{3121--3124},
  \doiprefix\url{10.1103/PhysRevLett.68.3121} (\bibinfo{year}{1992}).

\bibitem{huttner_quantum_1995}
\bibinfo{author}{Huttner, B.}, \bibinfo{author}{Imoto, N.},
  \bibinfo{author}{Gisin, N.} \& \bibinfo{author}{Mor, T.}
\newblock \bibinfo{journal}{\bibinfo{title}{Quantum cryptography with coherent
  states}}.
\newblock {\emph{\JournalTitle{Physical Review A}}}
  \textbf{\bibinfo{volume}{51}}, \bibinfo{pages}{1863--1869},
  \doiprefix\url{10.1103/PhysRevA.51.1863} (\bibinfo{year}{1995}).

\bibitem{kaushal_optical_2017}
\bibinfo{author}{Kaushal, H.} \& \bibinfo{author}{Kaddoum, G.}
\newblock \bibinfo{journal}{\bibinfo{title}{Optical communication in space:
  {Challenges} and mitigation techniques}}.
\newblock {\emph{\JournalTitle{IEEE communications surveys \& tutorials}}}
  \textbf{\bibinfo{volume}{19}}, \bibinfo{pages}{57--96}
  (\bibinfo{year}{2017}).

\bibitem{ke_experimental_2018}
\bibinfo{author}{Ke, X.}, \bibinfo{author}{Yang, S.} \& \bibinfo{author}{Wang,
  J.}
\newblock \bibinfo{title}{Experimental {Study} of {Free} {Space} {Coherent}
  {Optical} {Communication} on 1km}.
\newblock In \emph{\bibinfo{booktitle}{2018 10th {International} {Conference}
  on {Advanced} {Infocomm} {Technology} ({ICAIT})}}, \bibinfo{pages}{61--65},
  \doiprefix\url{10.1109/ICAIT.2018.8686537} (\bibinfo{year}{2018}).

\bibitem{vincent_extracting_2008}
\bibinfo{author}{Vincent, P.}, \bibinfo{author}{Larochelle, H.},
  \bibinfo{author}{Bengio, Y.} \& \bibinfo{author}{Manzagol, P.-A.}
\newblock \bibinfo{title}{Extracting and composing robust features with
  denoising autoencoders}.
\newblock In \emph{\bibinfo{booktitle}{Proceedings of the 25th international
  conference on {Machine} learning}}, \bibinfo{pages}{1096--1103}
  (\bibinfo{publisher}{ACM}, \bibinfo{year}{2008}).

\bibitem{lohani_use_2018}
\bibinfo{author}{Lohani, S.}, \bibinfo{author}{Knutson, E.~M.},
  \bibinfo{author}{O'Donnell, M.}, \bibinfo{author}{Huver, S.~D.} \&
  \bibinfo{author}{Glasser, R.~T.}
\newblock \bibinfo{journal}{\bibinfo{title}{On the use of deep neural networks
  in optical communications}}.
\newblock {\emph{\JournalTitle{Applied optics}}} \textbf{\bibinfo{volume}{57}},
  \bibinfo{pages}{4180--4190} (\bibinfo{year}{2018}).

\bibitem{sanchez-lengeling_inverse_2018}
\bibinfo{author}{Sanchez-Lengeling, B.} \& \bibinfo{author}{Aspuru-Guzik, A.}
\newblock \bibinfo{journal}{\bibinfo{title}{Inverse molecular design using
  machine learning: {Generative} models for matter engineering}}.
\newblock {\emph{\JournalTitle{Science}}} \textbf{\bibinfo{volume}{361}},
  \bibinfo{pages}{360--365} (\bibinfo{year}{2018}).

\bibitem{jang_ep-2092:_2018}
\bibinfo{author}{Jang, B.}, \bibinfo{author}{Chang, J.}, \bibinfo{author}{Park,
  A.} \& \bibinfo{author}{Wu, H.}
\newblock \bibinfo{journal}{\bibinfo{title}{{EP}-2092: {Generative} {Model} of
  {Functional} {RT}-{Plan} {Chest} {CT} for {Lung} {Cancer} {Patients} {Using}
  {Machine} {Learning}}}.
\newblock {\emph{\JournalTitle{Radiotherapy and Oncology}}}
  \textbf{\bibinfo{volume}{127}}, \bibinfo{pages}{S1149}
  (\bibinfo{year}{2018}).

\bibitem{li_machine_2018}
\bibinfo{author}{Li, Z.}, \bibinfo{author}{Meier, M.-A.},
  \bibinfo{author}{Hauksson, E.}, \bibinfo{author}{Zhan, Z.} \&
  \bibinfo{author}{Andrews, J.}
\newblock \bibinfo{journal}{\bibinfo{title}{Machine {Learning} {Seismic} {Wave}
  {Discrimination}: {Application} to {Earthquake} {Early} {Warning}}}.
\newblock {\emph{\JournalTitle{Geophysical Research Letters}}}
  (\bibinfo{year}{2018}).

\bibitem{donahue_exploring_2018}
\bibinfo{author}{Donahue, C.}, \bibinfo{author}{Li, B.} \&
  \bibinfo{author}{Prabhavalkar, R.}
\newblock \bibinfo{title}{Exploring speech enhancement with generative
  adversarial networks for robust speech recognition}.
\newblock In \emph{\bibinfo{booktitle}{2018 {IEEE} {International} {Conference}
  on {Acoustics}, {Speech} and {Signal} {Processing} ({ICASSP})}},
  \bibinfo{pages}{5024--5028} (\bibinfo{publisher}{IEEE},
  \bibinfo{year}{2018}).

\bibitem{torlai_neural-network_2018}
\bibinfo{author}{Torlai, G.} \emph{et~al.}
\newblock \bibinfo{journal}{\bibinfo{title}{Neural-network quantum state
  tomography}}.
\newblock {\emph{\JournalTitle{Nature Physics}}} \textbf{\bibinfo{volume}{14}},
  \bibinfo{pages}{447} (\bibinfo{year}{2018}).

\bibitem{gondara_medical_2016}
\bibinfo{author}{Gondara, L.}
\newblock \bibinfo{title}{Medical image denoising using convolutional denoising
  autoencoders}.
\newblock In \emph{\bibinfo{booktitle}{Data {Mining} {Workshops} ({ICDMW}),
  2016 {IEEE} 16th {International} {Conference} on}}, \bibinfo{pages}{241--246}
  (\bibinfo{publisher}{IEEE}, \bibinfo{year}{2016}).

\bibitem{fichou_powerful_2018}
\bibinfo{author}{Fichou, D.} \& \bibinfo{author}{Morlock, G.~E.}
\newblock \bibinfo{journal}{\bibinfo{title}{Powerful {Artificial} {Neural}
  {Network} for {Planar} {Chromatographic} {Image} {Evaluation}, {Shown} for
  {Denoising} and {Feature} {Extraction}}}.
\newblock {\emph{\JournalTitle{Analytical chemistry}}}
  \textbf{\bibinfo{volume}{90}}, \bibinfo{pages}{6984--6991}
  (\bibinfo{year}{2018}).

\bibitem{cheng_deep_2018}
\bibinfo{author}{Cheng, X.}, \bibinfo{author}{Zhang, L.} \&
  \bibinfo{author}{Zheng, Y.}
\newblock \bibinfo{journal}{\bibinfo{title}{Deep similarity learning for
  multimodal medical images}}.
\newblock {\emph{\JournalTitle{Computer Methods in Biomechanics and Biomedical
  Engineering: Imaging \& Visualization}}} \textbf{\bibinfo{volume}{6}},
  \bibinfo{pages}{248--252} (\bibinfo{year}{2018}).

\bibitem{wang_intelligent_2017}
\bibinfo{author}{Wang, D.} \emph{et~al.}
\newblock \bibinfo{journal}{\bibinfo{title}{Intelligent constellation diagram
  analyzer using convolutional neural network-based deep learning}}.
\newblock {\emph{\JournalTitle{Optics Express}}} \textbf{\bibinfo{volume}{25}},
  \bibinfo{pages}{17150--17166}, \doiprefix\url{10.1364/OE.25.017150}
  (\bibinfo{year}{2017}).

\bibitem{zhang_intelligent_2018}
\bibinfo{author}{Zhang, J.} \emph{et~al.}
\newblock \bibinfo{journal}{\bibinfo{title}{Intelligent adaptive coherent
  optical receiver based on convolutional neural network and clustering
  algorithm}}.
\newblock {\emph{\JournalTitle{Optics Express}}} \textbf{\bibinfo{volume}{26}},
  \bibinfo{pages}{18684--18698}, \doiprefix\url{10.1364/OE.26.018684}
  (\bibinfo{year}{2018}).

\bibitem{lohani_turbulence_2018}
\bibinfo{author}{Lohani, S.} \& \bibinfo{author}{Glasser, R.~T.}
\newblock \bibinfo{journal}{\bibinfo{title}{Turbulence correction with
  artificial neural networks}}.
\newblock {\emph{\JournalTitle{Optics letters}}} \textbf{\bibinfo{volume}{43}},
  \bibinfo{pages}{2611--2614} (\bibinfo{year}{2018}).

\bibitem{khan_joint_2017}
\bibinfo{author}{Khan, F.~N.} \emph{et~al.}
\newblock \bibinfo{journal}{\bibinfo{title}{Joint {OSNR} monitoring and
  modulation format identification in digital coherent receivers using deep
  neural networks}}.
\newblock {\emph{\JournalTitle{Optics Express}}} \textbf{\bibinfo{volume}{25}},
  \bibinfo{pages}{17767--17776}, \doiprefix\url{10.1364/OE.25.017767}
  (\bibinfo{year}{2017}).

\bibitem{fan_demodulator_2017}
\bibinfo{author}{Fan, M.} \& \bibinfo{author}{Wu, L.}
\newblock \bibinfo{title}{Demodulator based on deep belief networks in
  communication system}.
\newblock In \emph{\bibinfo{booktitle}{2017 {International} {Conference} on
  {Communication}, {Control}, {Computing} and {Electronics} {Engineering}
  ({ICCCCEE})}}, \bibinfo{pages}{1--5},
  \doiprefix\url{10.1109/ICCCCEE.2017.7867643} (\bibinfo{year}{2017}).

\bibitem{kulin_end--end_2017}
\bibinfo{author}{Kulin, M.}, \bibinfo{author}{Kazaz, T.},
  \bibinfo{author}{Moerman, I.} \& \bibinfo{author}{de~Poorter, E.}
\newblock \bibinfo{journal}{\bibinfo{title}{End-to-end {Learning} from
  {Spectrum} {Data}: {A} {Deep} {Learning} approach for {Wireless} {Signal}
  {Identification} in {Spectrum} {Monitoring} applications}}.
\newblock {\emph{\JournalTitle{arXiv:1712.03987 [cs]}}}
  (\bibinfo{year}{2017}).
\newblock \bibinfo{note}{ArXiv: 1712.03987}.

\bibitem{proakis_digital_2000}
\bibinfo{author}{Proakis, J.}
\newblock \emph{\bibinfo{title}{Digital {Communications}: 4th (fourth)
  edition}} (\bibinfo{publisher}{McGraw-Hill Companies, The},
  \bibinfo{year}{2000}).

\bibitem{noauthor_quantum_nodate-1}
\bibinfo{title}{Quantum {Detection} and {Estimation} {Theory}, {Volume} 123 -
  1st {Edition}}.

\bibitem{becerra_experimental_2013}
\bibinfo{author}{Becerra, F.~E.} \emph{et~al.}
\newblock \bibinfo{journal}{\bibinfo{title}{Experimental demonstration of a
  receiver beating the standard quantum limit for multiple nonorthogonal state
  discrimination}}.
\newblock {\emph{\JournalTitle{Nature Photonics}}}
  \textbf{\bibinfo{volume}{7}}, \bibinfo{pages}{147--152},
  \doiprefix\url{10.1038/nphoton.2012.316} (\bibinfo{year}{2013}).

\bibitem{izumi_displacement_2012}
\bibinfo{author}{Izumi, S.} \emph{et~al.}
\newblock \bibinfo{journal}{\bibinfo{title}{Displacement receiver for
  phase-shift-keyed coherent states}}.
\newblock {\emph{\JournalTitle{Physical Review A}}}
  \textbf{\bibinfo{volume}{86}}, \bibinfo{pages}{042328},
  \doiprefix\url{10.1103/PhysRevA.86.042328} (\bibinfo{year}{2012}).

\bibitem{martin_abadi_tensorflow:_2015}
\bibinfo{author}{{Martín Abadi}} \emph{et~al.}
\newblock \emph{\bibinfo{title}{{TensorFlow}: {Large}-{Scale} {Machine}
  {Learning} on {Heterogeneous} {Systems}}} (\bibinfo{year}{2015}).

\end{thebibliography}

 \section*{Acknowledgements}

This work was supported by the Office of Naval Research grant number N00014-19-1-2374.  Additionally, the research was supported in part using high performance computing (HPC) resources and services provided by Technology Services at Tulane University, New Orleans, LA.

\section*{Author contributions statement}
S.L. conceived and designed the neural network system, and ran all simulations.  R.T.G. developed and supervised the project. Both authors prepared the manuscript.

\end{document}